# The energy cost of protein messages lead to a new protein information law

Robersy Sánchez[1, 2] and Ricardo Grau[2]

[1] Research Institute of Tropical Roots, Tuber Crops and Banana (INIVIT). Biotechnology group. Santo Domingo. Villa Clara. Cuba.

[2] Center of Studies on Informatics. Central University of Las Villas. Villa Clara. Cuba.

**Abstract**

By considering the energy cost of messages carried by proteins as proportional to their information content we found experimental proof that proteins from all living organisms tend to have their estimated semantic content of information per unit mass, statistically, close to a constant. Thus, in the message carried by proteins –to achieve minimum energy waste– the rate of information content per unit mass tends to be optimized in living organisms. The experimental evidence of this new information law resembles a marathon where highly optimized proteins correspond to advanced runners followed by a main bunch and the stragglers –lowly optimized proteins. Our results suggest the existence of a continuous optimization process that living organisms had to face, in which a compromise between biological functionality, economic feasibility and the survival requirements is established.

## 1. Introduction

In information theory proteins are usually considered messages. A one-dimensional genetic message is recorded in a sequence of amino acids which fold up in a three-dimensional active protein [1, 2]. So their information content has been estimated by multiple researchers [1, 3]. Shannon in 1948 established information theory as a mathematical theory of communication [4]. In the second paragraph of his classical paper it is pointed out that: <u>Frequently the messages have meaning; that is they refer to or are correlated according to some system with certain physical or conceptual entities. These semantic aspects of communication are irrelevant to the engineering problem.</u> About this Yockey said: <u>Shannon is explaining that your telephone system can send or receive your message without having to understand a word you say because the sequence comprising the message need not have any meaning for the communication system. The communication system is just as successful when transmitting gibberish accurately between two points as it is when transmitting speeches from Shakespeare</u> [5]. Evidently, the semantic content of a message does not modify the telephone system's state. This is not, however, the situation found in proteins. A protein is not only considered a message, it is also a receiver device that can alter its state when its message is changed. Just a one "letter" change – an amino acid change– in a protein message may be sufficient to alter its structure and function (e.g. see the mutation databases in: [http://www.genomic.unimelb.edu.au/mdi/dblist/dblist.html](http://www.genomic.unimelb.edu.au/mdi/dblist/dblist.html)).

In addition, proteins also have been considered molecular machines and even molecular automata [6-10]. This means that, proteins are molecular devices where hardware and software function in concert. As a result, the software information value should be proportional to the hardware energetic cost. Following this claim, here we show that a new information concept, the value of protein information, allows us to express the semantic content of protein information [11], revealing a strong correlation between the protein molecular weight and the value of protein information.

**2. Theoretical Model**

Firstly, we devoted ourselves to finding a new approach to express the semantic content of protein information. Our starting point is the recently revealed Boolean structure of the genetic code [12, 13]. This Boolean structure led to a new point of view of the genetic information system as a Boolean information system. In this information system, the base sequences are written in an alphabet of four letters (four DNA bases) while the amino sequences are written in an alphabet of twenty letters or 20 amino acids [2]. Therefore, the DNA base triplets - codons- in the sequence can be similar to words and the synonymous codons coding for the same amino acid can act as words with the same meaning. From this point of view the amino acids become the meaning of the codons and we could then refer to the biological value of this meaning.

2.1. The value of protein information

Like Shannon we regarded the generation of a message to be a Markov process [4]. In particular, we considered the amino acid sequence as a message where every word is generated according to its deduction probability by a discrete information source represented by a first order Markov chain. This means that the deduction probability of the amino acid $a_k = i$ in the position $k$ depends only on the amino acid $a_{k-1} = j$ found in the previous position $k-1$ in the amino acid sequence. This is expressed by conditional probabilities $p(a_k = i \mid a_{k-1} = j)$. In addition, if we suppose that probabilities $p(a_k = i \mid a_{k-1} = j)$ are independent of the positions, i.e. if $p(a_k = i \mid a_{k-1} = j) = p(i \mid j)$ then, these are computed as:

$$p(i \mid j) = p(i, j)/p(j) = n(i, j)/n(j)$$

where $p(i, j)$ and $n(i, j)$ are the joint deduction probability and the number of joint deductions of amino acids $i$ and $j$ from the genetic code codons, respectively, $p(j)$ and $n(j)$ are the deduction probability and the total deductions for amino acid $j$ [14].

Next, the deduction probabilities $p(i \mid j)$ allow us to define a new concept, the value of protein information. Following Volkenshtein's original idea, the information value should express the measure of the non-substitutional character of information as well as the measure of its non-redundancy in the biological phylogenetic and ontogenetic development [11]. Amino acid deductions have a physicochemical meaning [13]. What's more the number of deductions is associated to the number of codons assigned to each amino acid (synonym quota) and the non-substitutional character of their information in the molecular evolution process. As a result, for each fixed amino acid the redundancy degree of its information is suggested by the number of its total deductions. Thus, the value of information eventually decreases with the increase of this number. This analysis led us to define the value of amino acid information in a previous paper [14]. Now, we define the value of protein information ($V(x)$) as:

$$V(x) = - \text{Log}_4 \, p_x$$

Where $p_x$ is the deduction probability of the protein $x$. According to our Markovian information source model, $p_x$ is computed as:

$$p_x = p(a_1) \, p(a_2 \mid a_1) \, p(a_3 \mid a_2) \ldots p(a_n \mid a_{n-1})$$

The choice of a logarithmic base corresponds to the choice of a unit for measuring the value of information. If base 4 is used, the resulting units may be called tetra-digits, or more briefly teds. If we want to express the value of information in bits, base-2 logarithm must be taken.

The protein information value should express the semantic content of the information, ignored by the classical information theory. The protein information value is not a measure of the biological value of protein for the cell. The semantic content of protein information has sense only to the receiver, i.e. to the protein itself.

2.2. Energy cost of the protein message and Value of protein information

Next, in terms of cellular economy a protein should carry just the maximum semantic content of information ($V(x)$) with minimum energy cost ($E(x)$). Since any molecular machine must dissipate at least $\varepsilon = kTLn(2)$ of energy (about $3 \times 10^{-21}$ Joule at room temperature) for each bit of information it erases or throws away [7, 15-17] the energy cost of information carried by a protein can be supposed to be equal to:

$$E(x) = \tfrac{1}{2}\, \varepsilon\, V(x) \quad (1)$$

At the same time, because the protein environment in the cell can be considered a thermal bath, the energy cost $E(x)$ should be equal to the maximum oscillation energy of the protein $x$ in a thermal bath. That is to say, equation (1) should state an energy limit for protein oscillation in a thermal bath. So we have:

$$E(x) = \tfrac{1}{2}\, MW(x)\, v^2_{max} \quad (2)$$

where $MW(x)$ is the molecular weight of the protein $x$ and $v_{max}$ is the maximum oscillation velocity reachable by any protein at temperature $T$. As a result, combining equation (1) and (2) we have:

$$V(x) = v^2_{max}/\varepsilon\, MW(x) \quad (3)$$

Explicitly, in any molecular machine, where "hardware" and "software" function as one, the semantic content of information is proportional to its molecular weight. At constant temperature, equation (3) is equivalent to the equality:

$$V(x)/MW(x) = v^2_{max}/\varepsilon = \text{constant} \quad (3')$$

## 3. Results

Next, we will see that equation (3) is supported by the experimental data. The classical ordinary least square method was applied to estimate the coefficients of a linear regression [18]:

$$V(x) = v^2_{max}/\varepsilon \; MW(x) + c + \delta \quad (4)$$

where $v^2_{max}/\varepsilon$ and intercept $c$ had to be estimated and $\delta$ was the model error. In the linear regression analysis firstly, the intercept $c$ was not considered significant ($p > 0.364$) –in equation (4)– and then, the final model was:

$$V(x) = v^2_{max}/\varepsilon \; MW(x) + \delta \quad (5)$$

The statistical model (5) corresponds to the theoretical equation (3). Figure 1 shows the linear regression analysis of $V(x)$ versus $MW(x)$ in a sample of 805 varied proteins sequences. The adjusted R square is 0.99986 and the proportional constant $v^2_{max}/\varepsilon$ is 19.56354547416 teds mol/kg ($MW(x)$ in units of kg/mol) or $2.35628690364 \times 10^{25}$ bits/kg. The regression hypothesis of the normal distributions of residuals was verified with the One-sample Kolmogorov-Smirnov normality test.

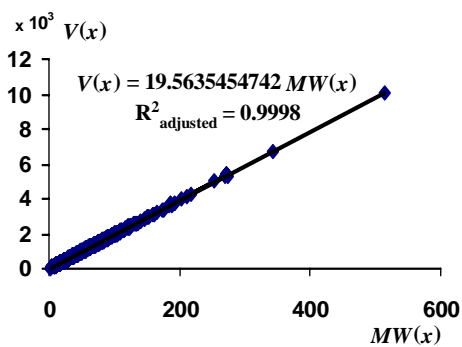

**Figure 1.** Regression of the value of protein information ($V(x)$) versus Molecular weight.($MW(x)$) for 805 proteins.

**Table 1.** Statistics of the linear regression through the origin of the $V(x)$ versus $MW(x)$ for 805 proteins [a].

| Unstandardized Coefficients | | t | Sig. | 95% Confidence Interval for B | | Adjusted R Square | Durbin-Watson |
|---|---|---|---|---|---|---|---|
| B | Std. Error | | | Lower Bound | Upper Bound | | |
| 19.56354547 | 0.00820966 | 2382.991 | 0.000 | 19.54743058 | 19.57966037 | 0.9998583 | 1.97618 |

| One-Sample Kolmogorov-Smirnov Test for Standardized Residuals | | |
|---|---|---|
| Kolmogorov-Smirnov Z | | 0.7415 |
| Asymp. Sig. (2-tailed) | | 0.6414 |
| Monte Carlo Sig. (2-tailed) | | 0.6283 |
| Sig.99% Confidence Interval | Lower Bound | 0.6159 |
| | Upper Bound | 0.6407 |

[a] The regression coefficient is highly significant ($p < 0.000$), there are not autocorrelation of residuals (Durbin-Watson coefficient is close to 2 and Box-Ljung $Q = 0.103$ with a $p > 0.748$ ) and their normal distribution is not rejected ($p > 0.5$). The model with an intercept has the constant not significant ($p > 0.364$).

The statistics applied for the analysis of the serial correlation in the residuals were Durbin-Watson d test and Box-Ljung Q statistic. All these statistics are presented in Table 1. In a starting sample of 956 proteins, for some of them the standardized regression residuals are greater than two standard deviations (they are outliers). However, the regression coefficient and the adjusted R square are not sensibly affected and keep their values close to the one reported. When we excluded these exceptional proteins in the regression analysis, we found –in remainder of 805 proteins– that residuals complied with the regression hypotheses (see Table 1).

Next, for a sample of 956 proteins the behavior of the rate $V(x)/MW(x)$ was analyzed by means of One-sample Kolmogorov-Smirnov normality test. For the Kolmogorov-Smirnov Z value the asymptotic and the Monte Carlo two-tailed signification were estimated. According to equation (3') it is expected that this rate should statistically equal to a constant.

We found deviations from the normal distribution produced by a small number of proteins with high and low rate values. Without these extreme values the remaining sub sample of 915 proteins followed a normal distribution with a mean of 19.5298805 teds mol/kg and 95% confidence interval: upper bound 19.5054295 and lower bound 19.5543309. Thus in this sample the rate was statically close to a constant according to what was expected in equation (3') (Table 2).

Besides this, a similar behavior was found in a sample 211 protein sequences of Cytochrome C from different living organisms but with a mean rate value of 20.327259064 teds mol/kg and 95% confidence interval: upper bound 20.285101445 and lower bound 20.369416721.

The specific effect of mutational events over the rate $V(x)/MW(x)$ –in a wild type gene– was also analyzed in a sample of 416 mutant DNA sequences of HIV-1 protease gene (previously translated to proteins).A similar result was observed with a mean rate value equal to 20.053628119 teds mol/kg and 95% confidence interval: upper bound 20.039997268 and lower bound 20.067258970 (see Table 2).

All samples of proteins and DNA sequences used in the statistical analyses were taken from the NCBI database at http://www.ncbi.nlm.nih.gov. The DNA sequences were translated into proteins to be used. All these sequences and the probabilities used to compute the value of protein information are available as supplementary material on the journal's web site.

**Table 2.** One-sample Kolmogorov-Smirnov normality test for the rate $V(x)/MW(x)$ for proteins and Cytochrome C from multiples living organisms and for HIV-1 protease [a].

|  |  | Proteins | Cytochrome C | HIV Protease |
|---|---|---|---|---|
| N |  | 915 | 211 | 416 |
|  | Median | 19.5287569 | 20.292384432 | 20.051671545 |
| Normal Parameters | Mean | 19.5298805 | 20.327259064 | 20.053628119 |
|  | Std. Deviation | 0.37685868 | 0.3106414676 | 0.1414337699 |
| Extreme Values | Minimum | 18.5204075 | 19.414553114 | 19.654524650 |
|  | Maximum | 20.4855354 | 21.368799403 | 20.483060580 |
| 95% Confidence Interval for Mean | Lower Bound | 19.5054295 | 20.285101445 | 20.039997268 |
|  | Upper Bound | 19.5543309 | 20.369416721 | 20.067258970 |
| Kolmogorov-Smirnov Z |  | 0.52883607 | 0.8712164163 | 0.4760935903 |
| Asymp. Sig. (2-tailed) | Sig. | 0.94246125 | 0.4336739182 | 0.9772173762 |
| Monte Carlo Sig. (2-tailed) | Sig. | 0.9412 | 0.4141 | 0.9716 |
| 99% Confidence Interval | Lower Bound | 0.93514036 | 0.4014123409 | 0.9673212190 |
|  | Upper Bound | 0.94725964 | 0.4267876591 | 0.9758787810 |

[a] The samples of proteins All these sequences are available as supplementary data on the journal's web site.

3.1 Stochastic Simulation

6422 arbitrary proteins were taken from the web site http://www.ncbi.nlm.nih.gov/entrez/ to compute the joint frequency matrix of amino acid pairs. From this matrix were obtained the marginal probability vector and a conditional probability matrix. Next, a simple Markov Chain Monte Carlo algorithm without acceptance probability restriction was used to generate 1900 random protein sequences. As we can see in Table 3 there is not statistical difference between the regression coefficients of this random sample (19.528827873) and the sample of 915 natural proteins (19.5287569) presented in Table 2.

**Table 3.** Statistics of the linear regression through the origin of the V(x) versus MW(x) for 1900 random protein sequences generated by means of a simple Markov Chain Monte Carlo algorithm without acceptance probability restriction.

| Unstandardized Coefficients | | t | Sig. | 95% Confidence Interval for B | | Adjusted R Square | Durbin-Watson |
|---|---|---|---|---|---|---|---|
| B | Std. Error | | | Lower Bound | Upper Bound | | |
| 19,528827873 | 0,0041832 | 4668,421 | 0.000 | 19,521 | 19,537 | 0,9999 | 1,992 |

## 4. Discussions

The correlation found is the manifestation of a new statistical protein information law. The law expresses the solution to an optimization process that living organisms had to face. In the molecular evolution process for the vast majority of proteins in living organism the ratio between the value of protein information and their mass ($V(x)/MW(x)$) tends to be statistically close to a constant value. In practice this phenomenon resembles a marathon where first are found advanced runners, next, the main bunch and finally, the stragglers. For instance, initially for 956 proteins we found deviations from a normal distribution produced by proteins with high and low values of the rate $V(x)/MW(x)$, "advanced runners" ($V(x)/MW(x) \geq 20.7$) and "stragglers" ($V(x)/MW(x) \leq 18.4$). The "leader runner" in this sample of proteins was the mouse Elastin precursor ($V(x)/MW(x) = 23.65.64$ teds mol/kg) and the Histidine operon leader peptide was the "straggler" ($V(x)/MW(x) = 17.4063$ teds mol/kg). Proteins with intermediate values of the $V(x)/MW(x)$ –a remaining of 915 proteins– have a remarkable normal distribution (see Table 2).

This result suggests an amazing question: Are proteins still evolving? Evidently, a protein with a high rate not necessarily belongs to a higher living organism. In particular this is reflected in the isofunctional family of Cytochrome C. In a sample of 211 enzymes –from different living organisms– the "leader runner" is the Cytocrome C from Rhodopseudomonas acidophila with a rate of 21.3688 teds mol/kg while the rate of the Homo sapiens Cytochrome C is 19.9771. Actually, it seems to be that for every family of isofunctional proteins there is a variation range of rate values following normal distribution (Table 2). A significant example is found in the sample of 416 DNA mutant sequences of HIV-1 protease gene isolated from different patients around the world. By reason of the technical limitations every mutational variant isolated from a patient is likely the most successful mutant of the virus population in this person, defeating the resistance of immune system and drug therapies, i.e., technically it is only possible isolate the most abundant mutant in a blood sample from a patient. As we see in Table 2 the success in the mentioned battle it is reached in a small variation range of rate values.  These examples suggest that the living organisms at the molecular level look for a compromise between biological functionality, economic feasibility and the survival requirements, in such away that the DNA polymorphism observed in a given gene is the result of an optimization process.

## 5. Conclusions

Given that proteins are molecular devices where hardware and software function as one in such a way that the software information value is proportional to the hardware energy cost, we found theoretical and experimental proof that proteins –from all living organisms– tend to have their estimated semantic content of information proportional to their molecular weight.

The semantic content of protein message is estimated on a new point of view of the genetic information system as a Boolean information system. Here, like Shannon we regarded the generation of a message to be a Markov process. As a result, in the message carried by proteins -for minimum energy waste- the rate of information content per unit mass tends to be statistically constant and close to $2.35628690364 \times 10^{25}$ bits/kg. The law expresses the solution to a continuous optimization process that living organisms had to face, in which there is a compromise between biological functionality, economic feasibility and the survival requirements.

**Acknowledgments**

This research was supported within the framework of a VLIR-IUS Collaboration Programme.

**References**

[1] H.P. Yockey, Information Theory and Molecular Biology (Cambridge Universty Press, Cambridge, UK, 1992)

[2] H. P. Yockey. Origin of life on earth and Shannon's theory of communication. Computer and Chemistry, 24 (2000) 105-123.

[3] O. Weiss, M.A. Jimenez-Montano, H..Herzel. Information content of protein sequences. .J Theor Biol. 206 (2000) 379-86.

[4] C.E. Shannon. A Mathematical Theory of Communication. Bell Syst. Tech. J., 27 (1948.) 379-423.

[5] H. P. Yockey. Information Theory, evolution and origin of life. Fundamentals of Life. Chapter II.10. (Editions scientifiques et médicales Elsevier SAS, 2002)

[6] T.D. Schneider. Theory of Molecular Machines. I. Channel Capacity of Molecular Machines. J. Theor. Biol., 148 (1991) 83-123.

[7] T.D. Schneider. Theory of Molecular Machines. II. Energy Dissipation from Molecular Machines. J. Theor. Biol. 148 (1991) 125-137.

[8] P.C. Marijuán, and J. Westley. Enzymes as molecular automata: a reflection on some numerical and philosophical aspects of the hypothesis. Biosystems, 27 (1992) 97-113.

[9] D. Bray. Protein molecules as computational elements in living cells. Nature, 376 (1995) 307-12.

[10] R. Lahoz-Beltra. Evolving hardware as model of enzyme evolution. Biosystems. 2001 Jun;61(1):15-25.

[11] M.V. Volkenshtein. Biophysics (Publishing House "Mir", Moscow, 1983)

[12] R. Sánchez, , R. Grau, and E. Morgado. The Genetic Code Boolean Lattice. MATCH Commun. Math. Comput. Chem., 52 (2004) 29-46.

[13] R. Sánchez, E. Morgado, R Grau. A genetic code boolean structure I. The meaning of boolean deductions. Bull. Math. Biol., 67 (2005) 1–14.

[14] R. Sánchez and R.Grau. A Genetic Code Boolean Structure. II. The Genetic Information System as a Boolean Information System. Bull. Math. Biol., doi:10.1016/j.bulm.2004.12.004 (2005)

[15] R. Landauer. Irreversibility and heat generation in the computing process. IBM Journal of Research and Development, 5 (1961) 183-191.

[16] R. Landauer. Minimal Energy Requirements in Communication, Science 272 (1996) 1914-1918.

[17] R. Landauer. Energy needed to send a bit. Proc. Royal Society of London, Series A 454 (1998) 305-311.

[18] R.B. Darlington. Regression and linear models (New York, McGraw-Hill, 1990).